\documentclass[preprint,eqsecnum,nofootinbib]{revtex4}
\usepackage{amssymb,amsbsy,bm,amsmath,psfrag,dsfont}
\usepackage{hyperref}
\usepackage{graphicx}
\usepackage{epsfig}
\usepackage[dvips]{color}

\begin{document}

\newcommand{\be}{\begin{equation}}
\newcommand{\ee}{\end{equation}}
\newcommand{\bea}{\begin{eqnarray}}
\newcommand{\eea}{\end{eqnarray}}
\newcommand{\no}{\noindent}

\newcommand{\la}{\lambda}
\newcommand{\si}{\sigma}
\newcommand{\vp}{\mathbf{p}}
\newcommand{\vk}{\vec{k}}
\newcommand{\vx}{\vec{x}}
\newcommand{\om}{\omega}
\newcommand{\Om}{\Omega}
\newcommand{\ga}{\gamma}
\newcommand{\Ga}{\Gamma}
\newcommand{\gaa}{\Gamma_a}
\newcommand{\al}{\alpha}
\newcommand{\ep}{\epsilon}
\newcommand{\app}{\approx}
\newcommand{\uvk}{\widehat{\bf{k}}}
\newcommand{\OM}{\overline{M}}

\title{Cosmological Particle Decays at Finite Temperature}
\author{Chiu Man Ho} \email{cmho@msu.edu}
   \affiliation{Department of
  Physics and Astronomy, Michigan State University, East Lansing, MI 48824, USA}
\author{Robert J. Scherrer} \email{robert.scherrer@vanderbilt.edu}
  \affiliation{Department of
  Physics and Astronomy, Vanderbilt University, Nashville, TN 37235, USA}
\date{\today}

\begin{abstract}
We calculate finite-temperature corrections to the decay rate of a generic neutral (pseudo)scalar particle that
decays into (pseudo)scalars or fermion-antifermion pairs. The ratio of the finite-temperature decay rate to the zero-temperature
decay rate is presented. Thermal effects are largest in the limit where the decaying particle is nonrelativistic but with a
mass well below the background temperature, but significant effects are possible even
when we relax the former assumption.  Thermal effects are reduced for the case of nonzero momentum
of the decaying particle. We discuss cosmological scenarios under which significant finite-temperature corrections to the decay rate
can be achieved.
\end{abstract}
\maketitle

\section{Introduction}

The cosmological consequences of particles decaying out of thermal equilibrium have long been a subject
of interest (see, e.g., the early work in Refs. \cite{DicusKolb,Lindley1,Weinberg1,ST1,Lindley2,ST2,ST3,Ellis}).
Nearly all studies of this kind have neglected the effect of finite (i.e., nonzero) temperatures on the decay rate.
This is often a reasonable approximation, depending on the parameters governing the decay.  However, a few authors
have examined thermal effects, with results that are scattered throughout the literature.  Weldon
\cite{weldon} provided one of the early treatments using finite-temperature field theory, the approach we use here.
Related calculations subsequently appeared in Refs. \cite{me,Yokoyama1,Yokoyama2,Arjun,Drewes1,Drewes2}.
Earlier discussions of corrections to the neutron decay rate relevant to primordial nucleosynthesis were given
in Refs. \cite{Dicus,Cambier}, and corrections to the Higgs decay rate into
electron-positron pairs can be found in Ref. \cite{Donoghue}, but these calculations use a different formalism.
Later Keil \cite{Keil1} and Keil and Kobes \cite{Keil2} reexamined the corrections to Higgs decay into $e^+e^-$ using the
real-time formulation of finite-temperature field theory. Related calculations for a scalar decaying into fermions were
done in Refs. \cite{Yokoyama1,Drewes2}. Recently, Gupta and Nayak \cite{Gupta} considered corrections to pseudoscalar decay into
two photons, and Czarnecki et al. \cite{CKLM} examined thermal corrections to the decay rate of charged fermions.

Here we provide a systematic calculation of finite-temperature corrections for neutral decaying particles.  Our goal
is to provide a more organized and systematic approach to this problem in a way that will be useful for researchers
in the future.  In the next section, we provide the formalism for our calculation.  In Sec. III, we examine three
cases of interest:  (A) a (pseudo)scalar decaying into (pseudo)scalars, (B) a pseudoscalar decaying into a fermion-antifermion
pair, and (C) a scalar decaying into a fermion-antifermion pair.  Case (A) and case (C) were examined previously
in Refs. \cite{weldon,Drewes2} and \cite{Keil1,Keil2}, respectively, but in neither case was the enhancement/suppression ratio to
the decay rate explicitly studied. Refs. \cite{Keil1,Keil2} considered a scalar that decays at zero momentum, while our results for
case (C) are valid for arbitrary momentum for the decaying particle. Case (B) has not been previously discussed in the literature.
In Sec. IV, we discuss our results and indicate the cosmological scenarios to which they are applicable.  The most striking effect
is the possible enhancement of the decay rate for the case of decays into (pseudo)scalars.
As we show in Sec. IV, an extremely large enhancement is difficult (except for reheating after inflation),
but not impossible to achieve in the context of the standard cosmological model. We also note that thermal corrections
are reduced as the momentum of the decaying particle increases, and we provide an explanation for this effect.

\section{Decay Rates at Finite Temperature}

At zero-temperature, the decay rate $\gamma_D$ of a particle with energy $E_0$ can be calculated by the Cutkosky rules \cite{Cutkosky}.
This leads to
\bea
\label{optical}
\gamma_D = -\frac{\textrm{Im}\, \Sigma_{T=0} (E_0)}{E_0}\,,
\eea
which relates the decay rate to the imaginary part of the self-energy $\textrm{Im}\, \Sigma_{T=0} (E_0)$ of the decaying particle and
its energy $E_0$.

At finite temperature $T$, the Cutkosky rules need to be modified. Using the imaginary-time formalism \cite{kapusta,lebellac},
Weldon \cite{weldon} showed that for a decaying particle with energy $E$ in the thermal bath, Eq. \eqref{optical} is modified into
\bea
\label{opticalT}
\Gamma_D \,\pm\, \Gamma_I = -\frac{\textrm{Im}\, \Sigma (E)}{E}\,,
\eea
where $\Gamma_D$ is the finite-temperature decay rate,
``+" and ``--" correspond to a decaying fermion and boson respectively, and $\Gamma_I$ is the inverse decay rate of the particles
resulting from the decay. Up to one-loop calculation, this result was confirmed by Kobes and Semenoff \cite{Kobes} who used the real-time
formulation \cite{kapusta,lebellac}. If the unstable particle decays in a thermal bath that is abundant in its decay products, the decay products
would have the probability to recombine in the thermal bath, and $\textrm{Im}\, \Sigma (E)$ accounts for both of the decay and recombination processes.

Weldon \cite{weldon} also showed that regardless of whether the decaying particle is a fermion or boson, the ratio of $\Gamma_D$ to $\Gamma_I$ is a universal function of $E$, namely
\bea
\frac{\Gamma_D}{\Gamma_I}=\exp{(\,\beta\,E\,)}\,,
\eea
with $\beta =1/T$. This allows us to derive the decay rate at finite temperature
\bea
\label{GammaD}
\Gamma_D = \frac{1}{1\,\pm\, e^{-\beta\,E}}\, \left(\,-\frac{\textrm{Im}\, \Sigma (E)}{E}\,\right)\,,
\eea
where again ``+" and ``--" correspond to a decaying fermion and boson respectively.

In this paper, we are interested in an unstable particle that is out of equilibrium.
We assume that the finite-temperature corrections to the mass of the decaying particle are negligible compared to its mass in the vacuum. So we can approximate $E$ as $E_0$. As we shall see, the imaginary part of the self-energy of the decaying particle can generally be written as a linear combination of zero-temperature and finite-temperature contributions, and with the approximation $E \approx E_0$, we can write
$\textrm{Im}\, \Sigma (E) \approx \textrm{Im}\, \Sigma_{T=0} ( E_0) + \textrm{Im}\, \Sigma_{T\neq0} ( E_0)$.
We can then define the ratio
\bea
\label{ratio}
R \,\equiv\, \frac{\Gamma_D}{\gamma_D}\,,
\eea
which characterizes the missing factor we would encounter if we blindly use the zero-temperature decay rate $\gamma_D$ in a
thermal bath. The calculation of R (generalized to arbitrary momentum for the decaying particle) for the cases of interest is the main goal of this paper (and the results that extend this work beyond that of \cite{weldon,Drewes2,Keil1,Keil2}).

\section{Specific Particle Decay Rates at Finite Temperature}

We illustrate our study by considering three simple models:
(A) a (pseudo)scalar decaying into (pseudo)scalars, (B) a pseudoscalar decaying into a fermion-antifermion
pair, and (C) a scalar decaying into a fermion-antifermion pair. In particular, we study the ratio $R \,\equiv\, \frac{\Gamma_D}{\gamma_D}$ and investigate how it changes with temperature. All of the calculations are done under the imaginary-time formalism. This formalism has the advantage that perturbation theory can still be organized into a diagrammatic expansion with the same vertices as at zero temperature.

\subsection{(Pseudo)scalar Decaying into (pseudo)scalars}

\begin{figure}[t!]
\includegraphics[height=3cm, width=5.4cm]{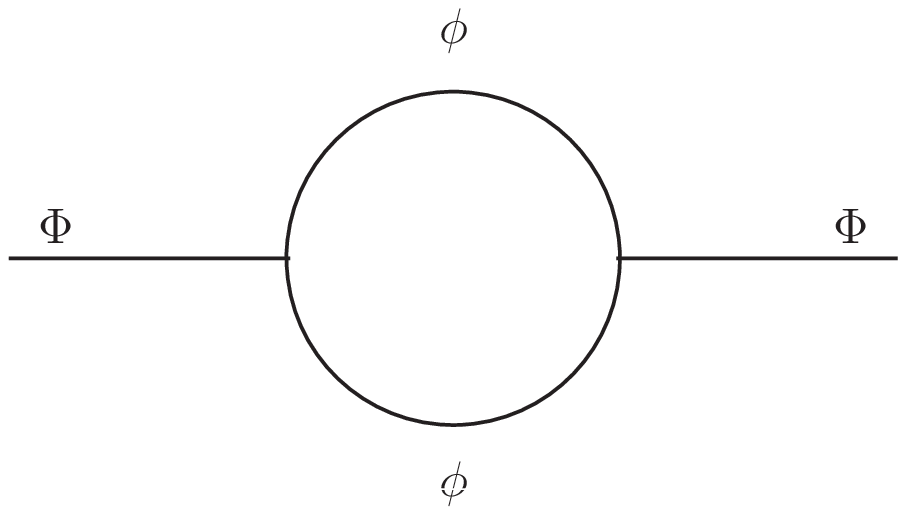}
~~~~~~~
\includegraphics[height=3cm, width=5.4cm]{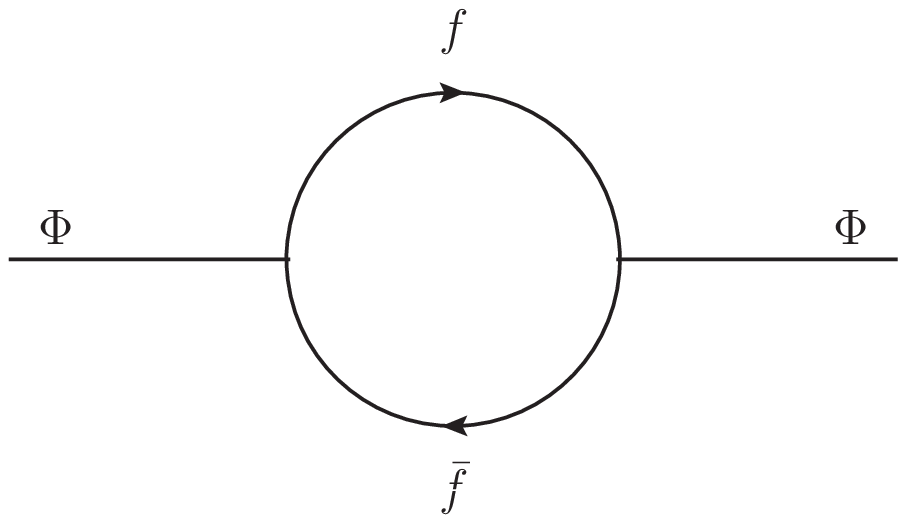}
\caption{(Left) Self-energy for $\Phi$ with a (pseudo)scalar $\phi$ loop.~ (Right) Self-energy for $\Phi$ with a fermion $f$ loop. }
\label{fig:self}
\end{figure}

We consider the model in which a (pseudo)scalar $\Phi$ can decay into a pair of identical (pseudo)scalars $\phi\,\phi$.
The interaction operator responsible for this process is:
\bea
\mathcal{L}_{\textrm{int}} = g\, \Phi\,\phi^2\,.
\eea
This model is relevant to several cases of interest. For instance, $\Phi$ could be
the Standard Model (SM) Higgs decaying into a pair of scalar dark matter particles \cite{ScalarSinglet},
or conversely, a scalar dark matter particle decaying into a pair of SM Higgs.
Alternately, $\Phi$ could be the SM Higgs
decaying into a pair of light CP-odd scalars in the Next-to-Minimal Supersymmetric Standard Model (NMSSM) \cite{NMSSM}, or
the heavier CP-odd Higgs decaying into two lighter CP-even Higgs in the two-Higgs-doublet-model (2HDM) \cite{Gunion}.
Finally, $\Phi$ could be the SM Higgs decaying into a pair of pseudo-goldstone bosons, which has been proposed by Weinberg
\cite{Weinberg2} to explain the fractional effective number of neutrinos hinted by Ref. \cite{Planck}.

To apply the Cutkosky rules, we need to calculate the self-energy of $\Phi$ as shown in Fig. \ref{fig:self} (left) and then put the $\phi$ particles on their mass-shell. Based on the calculations (of the imaginary part of the self-energy) in Appendix A1, Eq. \eqref{GammaD} and Eq. \eqref{optical}, we obtain the rates for the decay $\Phi \rightarrow \phi\,\phi$ at both zero and finite temperatures:
\bea
\gamma_D &=& \frac{g^{2}}{8\, \pi\, E_0}\,\sqrt{\,1-\frac{4\, m_{\phi}^{2}}{M_\Phi^2}} \;\Theta[\,M_\Phi^2-4 \,m_\phi^2\,]\,, \\
\Gamma_D &\approx&  \frac{1}{1\,-\, e^{-E_0/T}}\,\left[\, \gamma_D + \frac{g^{2}\,T}{4\, \pi\, E_0\,k}\,
\ln \left(\,
\frac{ 1-e^{-\omega^{+}/T} }{ 1-e^{-\omega^{-}/T}}\,\right) \;\Theta[\,M_\Phi^2-4 \,m_\phi^2\,]\,\right]\,,
\eea
\bea
R_{\Phi \rightarrow \phi\phi} \approx \,\frac{1}{1\,-\, e^{-E_0/T}}\,\left[\,1+ \frac{2\,T}{k\,\sqrt{\,1-\frac{4\, m_{\phi}^{2}}{M_\Phi^2}}}\,\ln \left(\,
\frac{ 1-e^{-\omega^{+}/T} }{ 1-e^{-\omega^{-}/T}}\,\right)\,\right] \,\Theta[\,M_\Phi^2-4 \,m_\phi^2\,]\,,
\eea
where $\Theta$ is the Heaviside step function and
\bea
E_0 = \sqrt{k^2+M_\Phi^2},~~~~~~\omega^{\pm}=\frac{E_0\,\pm\, k\,\sqrt{\,1-\frac{4\, m_{\phi}^{2}}{M_\Phi^2}}}{2}\,.
\eea
with $k$ being the momentum of the $\Phi$ particle.

Typically, $m_\phi$ would receive finite-temperature corrections which go like $\xi\, T$ where $\xi$ is a perturbatively small constant. For $\xi \lesssim 0.01 $ and $T \lesssim 50 M_\Phi$, the finite-temperature corrections to $m_\phi$ are negligible compared to $M_\Phi$ and therefore can be ignored in the quantity $m_\phi^2/M_\Phi^2$.

For the calculations in Appendix A1, we have used the thermal propagators for the $\phi$ particles and so they are required to be in thermal
equilibrium with the thermal bath. It is precisely these thermalized $\phi$ particles that induce finite-temperature corrections to the decay rate
of $\Phi \rightarrow \phi\,\phi$. To ensure that there is a significant abundance of the $\phi$ particles in the thermal bath, we also require that $T > m_\phi$.

\begin{figure}
\includegraphics[height=6cm, width=9cm]{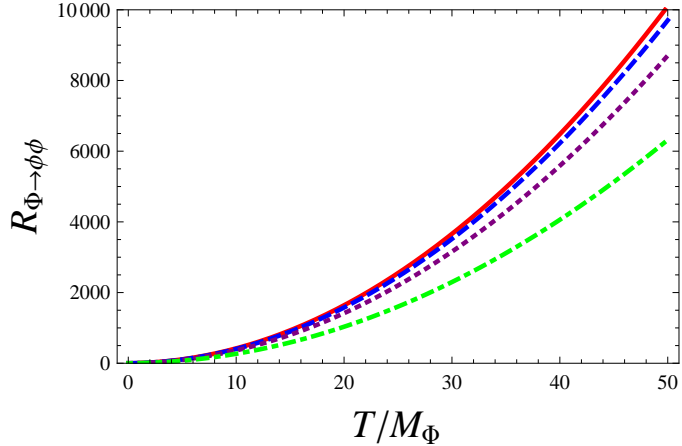}
\caption{The plot of $R_{\Phi \rightarrow \phi\phi}$ against $T/M_\Phi$ for a nonrelativistic $\Phi$, assuming $4\,m_\phi^2 \ll M_\Phi^2$.
The solid (red), dashed (blue), dotted (purple) and dot-dashed (green) lines correspond to the parameters $k/M_\Phi = 0.001,\, 0.25,\,0.5,\,1$ respectively.}
\label{fig:Rphi}
\end{figure}

\begin{figure}
\includegraphics[height=6cm, width=9cm]{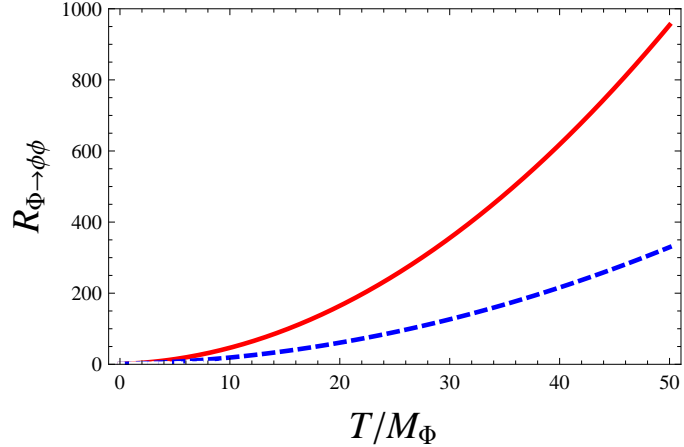}
\caption{The plot of $R_{\Phi \rightarrow \phi\phi}$ against $T/M_\Phi$ for a slightly relativistic $\Phi$, assuming $4\,m_\phi^2 \ll M_\Phi^2$.
The solid (red) and dashed (blue) lines correspond to the parameters $k/M_\Phi = 5,\, 10$ respectively.}
\label{fig:Rphi1}
\end{figure}

\begin{figure}
\includegraphics[height=6cm, width=9cm]{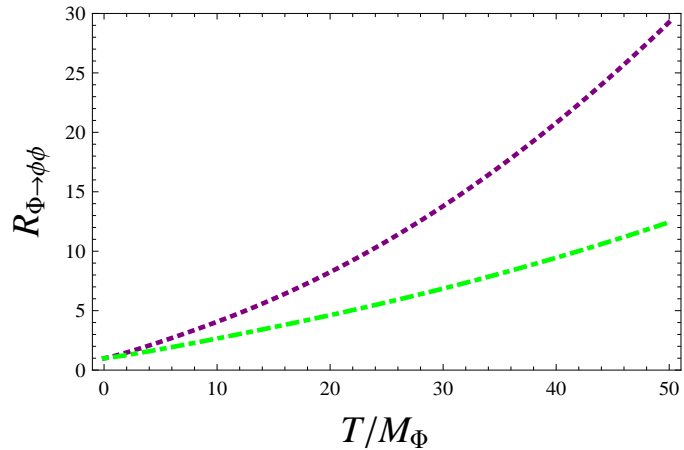}
\caption{The plot of $R_{\Phi \rightarrow \phi\phi}$ against $T/M_\Phi$ for a highly relativistic $\Phi$, assuming $4\,m_\phi^2 \ll M_\Phi^2$.
The dotted (purple) and dot-dashed (green) lines correspond to the parameters $k/M_\Phi = 50,\,100 $ respectively.}
\label{fig:Rphi2}
\end{figure}

Now consider an out-of-equilibrium (pseudo)scalar $\Phi$ which decays into a pair of
identical (pseudo)scalars $\phi\,\phi$. We plot $R_{\Phi \rightarrow \phi\phi}$ as a function of $T/M_\Phi$ in Fig. \ref{fig:Rphi}, Fig. \ref{fig:Rphi1} and Fig. \ref{fig:Rphi2}, taking $4\,m_\phi^2 \ll M_\Phi^2$.  This last assumption is not essential and is only used to simplify the plots; we have verified that
relaxing this assumption gives similar results.

It is clear that for a nonrelativistic $\Phi$ with $k/M_\Phi \lesssim 1$, the enhancement factor $R_{\Phi \rightarrow \phi\phi}$ can be as large as $10^4$ (see Fig. \ref{fig:Rphi}). For a slightly relativistic $\Phi$, the enhancement factor can be at least $10^2$ for $T \gtrsim 20 M_\Phi$ and can reach $10^3$ for $T \sim 50 M_\Phi$  (see Fig. \ref{fig:Rphi1}). Even for a highly relativistic $\Phi$, the enhancement factor can still reach 10 or higher for $T \sim 50 M_\Phi$ (see Fig. \ref{fig:Rphi2}). We thus conclude that the condition $T/M_\Phi \gg 1$ is the key for large thermal effects on the decay rate. The magnitude of the thermal effects increases significantly when we move from the relativistic limit to the nonrelativistic limit.

In the discussions above, we have considered $T$ as large as 50$M_\Phi$. We have not taken into account the possible finite-temperature correction
to $M_{\Phi}^2$. With a more accurate calculation, $M_{\Phi}^2$ at finite temperature $T$ should take the form
\bea
M_\Phi^2(T) = M_\Phi^2 + \Delta M_\Phi^2(T)\,,
\eea
where $\Delta M_\Phi^2(T)$ arises from the real part of the self-energy for $\Phi$. The quantity $\Delta M_\Phi^2(T)$
represents the finite-temperature correction to $M_\Phi^2$, and it is computed in Appendix A2:\footnote{Notice that in the limit
$m_\phi \rightarrow 0$, the quantity in Eq. \eqref{dalpha} becomes divergent. This is a manifestation of the infrared divergence due to
massless particles at finite temperature. The approximation made in Eq. \eqref{error} may no longer be consistent. In this case, a
simple analytical form for $\Delta M_\Phi^2(T)$ may not be available.}
\bea
\label{DeltaM}
\Delta M_\Phi^2(T) \approx \frac{g^{2}}{24}\,\,\frac{\omega^2- k^2}{\omega^2}\,\frac{T^2}{m_\phi^2}\,.
\eea
Note that the real part of the self-energy for $\Phi$, from which $\Delta M_\Phi^2(T)$ is extracted, was not computed in Ref. \cite{weldon} but discussed in Refs. \cite{Drewes2,DrewesPLB}.

For a nonrelativistic $\Phi$ with $k/M_\Phi \lesssim 1$, we have $\omega^2- k^2 \approx \omega^2 \approx M_\Phi^2$,
and so $\Delta M_\Phi^2(T) \approx \frac{g^{2}}{24}\,\frac{T^2}{m_\phi^2}$.  For consistency, we require $\Delta M_\Phi^2(T) \lesssim M_\Phi^2$.
In order to obtain $R_{\Phi \rightarrow \phi\phi} \gg 1$ in Fig. \ref{fig:Rphi}, we have taken $M_\Phi \lesssim T/5$, which gives:
\bea
\label{consistency}
\frac{g^{2}}{24}\,\frac{T^2}{m_\phi^2} \lesssim \left(\,\frac{T}{5}\,\right)^2  ~~~\Leftrightarrow ~~~ |g| \,\lesssim  \, m_\phi\,.
\eea
Therefore, $|g| \,\lesssim  \, m_\phi$ is the consistency condition that allows one to neglect the finite-temperature correction to $M_\Phi^2$ for
the range of $T/M_\phi$ values shown in Fig. \ref{fig:Rphi}.

In contrast, for a relativistic $\Phi$ with $k/M_\Phi \gtrsim 1$, we have $\omega^2- k^2 \approx  0$, and so $\Delta M_\Phi^2(T) \approx 0$. Thus, one can always neglect the finite-temperature correction to $M_\Phi^2$ for the range of $T/M_\phi$ values shown in
Fig. \ref{fig:Rphi1} and Fig. \ref{fig:Rphi2}. There is no analogous upper bound on $|g|$.

A final remark is in order. When the temperature is sufficiently higher than the masses of the particles under consideration,
naive perturbation theory may break down, especially for soft external momenta ($k \ll T$). One could correct this by using
a resummed perturbative expansion \cite{Braaten}. See Ref. \cite{Drewes2} for recent work that discusses this issue.

\subsection{Pseudoscalar Decaying into a Fermion-Antifermion Pair}

We consider the model in which a pseudoscalar $\Phi$ can decay into a fermion-antifermion pair $f\,\bar{f}$.
The interaction operator responsible for this process is:
\bea
\mathcal{L}_{\textrm{int}} = i\,\lambda \, \bar{f}\,\gamma^5\,\Phi\,f\,.
\eea
This model could be relevant in several cases. For instance, $\Phi$ could be a pseudoscalar dark matter candidate,
which is the neutral component of an SU(2) multiplet, decaying into SM fermions \cite{SU(2)DM}. Alternately, $\Phi$ could be the Majoron decaying into Majorana neutrinos \cite{Majoron}.

To apply the Cutkosky rules, we need to calculate the self-energy of $\Phi$ as shown in Fig. \ref{fig:self} (right) and then put the fermions $f$ and $\bar{f}$ on their mass-shell. Based on the calculations in Appendix B, Eq. \eqref{GammaD} and Eq. \eqref{optical}, we obtain the rates for the decay $\Phi \rightarrow f\,\bar{f}$ at both zero and finite temperatures:
\bea
\gamma_D &=& \frac{\lambda^{2}\,M_\Phi^2}{8\, \pi\, E_0}\,\sqrt{\,1-\frac{4\, m_{f}^{2}}{M_\Phi^2}} \;\Theta[\,M_\Phi^2-4 \,m_f^2\,]\,, \\
\Gamma_D &\approx&  \frac{1}{1\,-\, e^{-E_0/T}}\,\left[\, \gamma_D + \frac{\lambda^{2}\,M_\Phi^2\,T}{4\, \pi\,E_0\,k}\,
\ln \left(\,
\frac{ 1+e^{-E^{+}/T} }{ 1+e^{-E^{-}/T}}\,\right) \;\Theta[\,M_\Phi^2-4 \,m_f^2\,]\,\right]\,,
\eea
\bea
R_{\Phi \rightarrow f \bar{f}} \approx \,\frac{1}{1\,-\, e^{-E_0/T}}\,\left[\,1+ \frac{2\,T}{k\,\sqrt{\,1-\frac{4\, m_{f}^{2}}{M_\Phi^2}}}\,\ln \left(\,
\frac{ 1+e^{-E^{+}/T} }{ 1+e^{-E^{-}/T}}\,\right)\,\right] \,\Theta[\,M_\Phi^2-4 \,m_f^2\,]\,,
\eea
where
\bea
\label{Eplusminus}
E^{\pm}=\frac{E_0\,\pm\, k\,\sqrt{\,1-\frac{4\, m_{f}^{2}}{M_\Phi^2}}}{2}\,.
\eea

Similarly, $m_f$ would receive finite-temperature corrections which go like $\xi'\, T$ where $\xi'$ is a perturbatively
small constant.
For $\xi' \lesssim 0.01 $ and $T \lesssim 50 M_\Phi$, the finite-temperature corrections to $m_f$ are negligible compared
to $M_\Phi$ and therefore can be ignored in the quantity $m_f^2/M_\Phi^2$.

For the calculations in Appendix B, we have used the thermal propagators for the fermions $f$ and $\bar{f}$ and so they are required to be in thermal
equilibrium with the thermal bath. It is precisely these thermalized fermions $f$ and $\bar{f}$ that induce finite-temperature corrections to the decay rate of $\Phi \rightarrow f\,\bar{f}$. To ensure that there is a significant abundance of the fermions $f$ and $\bar{f}$ in the thermal bath, we also require that $T > m_f$.

\begin{figure}
\includegraphics[height=6cm, width=9cm]{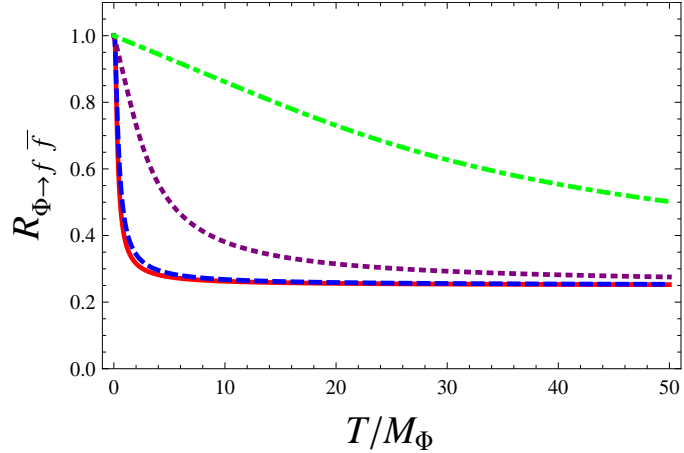}
\caption{The plot of $R_{\Phi \rightarrow f\bar{f}}$ against $T/M_\Phi$, assuming $4\,m_f^2 \ll M_\Phi^2$.
The solid (red), dashed (blue), dotted (purple) and dot-dashed (green) lines correspond to the parameters $k/M_\Phi = 0.001,\, 1,\,10,\,100$ respectively.}
\label{fig:Rf}
\end{figure}

Now consider an out-of-equilibrium pseudoscalar $\Phi$ that decays into a fermion-antifermion pair, $f\,\bar{f}$. We
plot $R_{\Phi \rightarrow f\bar{f}}$ against $T/M_\Phi$ in Fig. \ref{fig:Rf}.  As we can see, the suppression factor $R_{\Phi \rightarrow f\bar{f}}$ does not vary much when $T/M_\Phi$ increases from 1 to 50. When one moves from the nonrelativistic limit to the relativistic limit, the thermal effects decrease and hence the suppression factor increases (less suppressed).


\subsection{Scalar Decaying into a Fermion-Antifermion Pair}

We consider the model in which a scalar $\Phi$ can decay into a fermion-antifermion pair, $f\,\bar{f}$.
The interaction operator responsible for this process is:
\bea
\mathcal{L}_{f} = y \, \bar{f}\,\Phi\,f\,.
\eea
For instance, $\Phi$ could be a scalar dark matter candidate, which is the neutral component of an SU(2) multiplet, decaying into SM fermions \cite{SU(2)DM}. Besides, $\Phi$ could be the SM Higgs decaying into SM fermions.


As in the pseudoscalar case, in order to apply the Cutkosky rules, we need to calculate the self-energy of $\Phi$ as shown in Fig. \ref{fig:self}(right) and then put the fermions $f$ and $\bar{f}$ on their mass-shell. Based on the calculations in Appendix C, Eq. \eqref{GammaD} and Eq. \eqref{optical}, we obtain the rates for the decay $\Phi \rightarrow f\,\bar{f}$ at both zero and finite temperatures:
\bea
\gamma_D &=& \frac{y^{2}\,M_\Phi^2}{8\, \pi\, E_0}\,\left(\,1-\frac{4\, m_{f}^{2}}{M_\Phi^2}\,\right)^{3/2} \;\Theta[\,M_\Phi^2-4 \,m_f^2\,]\,, \\
\Gamma_D &\approx&  \frac{1}{1\,-\, e^{-E_0/T}}\,\left[\, \gamma_D + \frac{y^{2}\,M_\Phi^2\,T}{4\, \pi\,E_0\,k}\,
\left(\,1-\frac{4\, m_{f}^{2}}{M_\Phi^2}\,\right)\,\ln \left(\,
\frac{ 1+e^{-E^{+}/T} }{ 1+e^{-E^{-}/T}}\,\right) \;\Theta[\,M_\Phi^2-4 \,m_f^2\,]\,\right]\,, \nonumber \\
\eea
\bea
R'_{\Phi \rightarrow f \bar{f}} \approx \,\frac{1}{1\,-\, e^{-E_0/T}}\,\left[\,1+ \frac{2\,T}{k\,\sqrt{\,1-\frac{4\, m_{f}^{2}}{M_\Phi^2}}}\,\ln \left(\,
\frac{ 1+e^{-E^{+}/T} }{ 1+e^{-E^{-}/T}}\,\right)\,\right] \,\Theta[\,M_\Phi^2-4 \,m_f^2\,]\,,
\eea
where $E^{\pm}$ is given by \eqref{Eplusminus}.
Again, $m_f$ would receive finite-temperature corrections which go like $\xi'' \,T$ where $\xi''$ is a perturbatively
small constant.
For $\xi'' \lesssim 0.01 $ and $T \lesssim 50 M_\Phi$, the finite-temperature corrections to $m_f$
are negligible compared to $M_\Phi$ and therefore can be ignored in the quantity $m_f^2/M_\Phi^2$.
Moreover, we require that $f,\bar{f}$ are thermalized and $T > m_f$ for reasons explained in the pseudoscalar case.

We find that $R'_{\Phi \rightarrow f \bar{f}}$ is identical to $R_{\Phi \rightarrow f \bar{f}}$ and so we can just refer to Fig. \ref{fig:Rf}
for the behavior of $R'_{\Phi \rightarrow f \bar{f}}$ against $T/M_\Phi$. The conclusion is similar to the pseudoscalar case.


\section{Discussion and Cosmological Implications}

The results presented here are in broad agreement with our intuition from statistical mechanics.  For decays into fermions
(III.B. and III.C.), the result of finite-temperature effects is a suppression of the decay rate
for $T/M_\Phi \gg 1$ resulting from Pauli blocking.  Conversely, for decays into (pseudo)scalars (III.A.), one sees significant
enhancement from stimulated decays when $T/M_\Phi \gg 1$.

Our results show that thermal corrections are reduced for nonzero momentum of the decaying particle; in all of the scenarios we explored,
the finite-temperature effects decrease as $k/M_\Phi$ increases.  Note that this is {\it not} due to Lorentz suppression of the decay
rate, as the ratio $R$ defined in Eq. (\ref{ratio}) includes the same Lorentz factor in both the numerator and the denominator.  To understand this effect, suppose that $k/M_\Phi$ is large, and transform to the rest frame of the decaying particle.  In this frame, the thermal background
has a large net nonzero mean momentum.  But the particles in the thermal bath that interact to produce a reverse decay must have zero
total momentum, a result that becomes more difficult to achieve as the thermal background is boosted to higher and higher momentum.

When are these results relevant for cosmology?  The only cosmologically-relevant decay process known to occur with certainty
is the decay of free neutrons into protons during primordial nucleosynthesis, which occurs at a temperature $T \sim$ 0.1 MeV.
In this case, $T/M_{\textrm{neutron}} \ll 1$, so we expect thermal corrections to be very small, as they indeed are \cite{Dicus,Cambier}.

Now consider more hypothetical scenarios.  As we have seen, a large change in the decay rate occurs only for
$T/M_\Phi \gg 1$.  For a particle with a standard thermal history that drops out of equilibrium when it is nonrelativistic,
we automatically have $T/M_\Phi \ll 1$, so if this particle subsequently decays, thermal corrections to the decay
rate will be negligible (for this and other scenarios discussed here, see, e.g, Ref. \cite{KolbTurner}).

On the other hand, if the particle drops out of equilibrium while still relativistic, we would have $T/M_\Phi \gg 1$
when this decoupling occurs.  However, in this case $k \sim T$ at all later times,
so that $k /M_\Phi \sim T/M_\Phi \gg 1$.  Thus, if
the particle decays when $T/M_\Phi \gg 1$, it is still relativistic at decay ($k /M_\Phi \gg 1$).  In this scenario,
decay into (pseudo)scalars can still produce an enhancement of $O(10)$ (see Fig. 4), but not the $O(10^3)$ enhancement
in Fig. 2.  To achieve the latter requires the transfer of entropy into the thermal background so that $T/M_\Phi \gg 1$ when $k/M_\Phi \ll 1$.
Some entropy transfer occurs in the standard cosmological model when particles that are in thermal equilibrium become nonrelativistic
\cite{KolbTurner}.  A larger effect can occur in nonstandard scenarios
when nonrelativistic particles come to dominate the energy density of the thermal background and then decay out of equilibrium \cite{ST1}.  In either case, the thermal background will be heated so that $k < T$, and one could then have a decaying particle with $T/M_\Phi \gg 1$ and $k/M_\Phi \ll 1$.  (This loophole is in principle possible even when $\Phi$ decouples while nonrelativistic, but it would require
an enormous entropy release in this case).

A third possibility is a nonthermal production mechanism for the decaying particle in question.  For example, axions (or axion-like particles)
produced by the misalignment mechanism are ``born" with $T/M_a \gg 1$ and $k/M_a \ll 1$.

One possible cosmological scenario for which thermal effects might be significant is reheating after inflation. Once the inflatons start to decay, the decaying products may form a dense plasma which back-reacts on the inflaton decay \cite{Kolb}. It is possible that $T/M_\Phi \gg 1$ is satisfied during this period and our results apply. The effect of this dense plasma on the thermal history of the universe was investigated in Ref. \cite{Drewes3}. Our results may also be relevant to the fate of flat directions after reheating. It was pointed out that thermal corrections
could be significant in this context \cite{Yokoyama1,Buchmuller}.

Thus, while the conditions necessary for thermal corrections to produce an extremely large change in the decay rate are
somewhat unusual (except for reheating after inflation), they are not impossible to achieve in the context of our current
cosmological model.  Of course, our results are also valid in the case of smaller corrections to the decay rate, which are easier to achieve.

\bigskip

\acknowledgments
We thank Marc Kamionkowski for useful comments. C.M.H. was supported in part by the Office of the Vice-President for Research and Graduate
Studies at Michigan State University.  R.J.S. was supported in part by the Department of Energy (DE-SC0011981).

\appendix

\section{~~ $\Phi \rightarrow \phi \,\phi$}

\subsection{Decay Rate: ~ Imaginary Part of the Self-Energy}

The treatment in this subsection is similar to that of \cite{me}. The one-loop self-energy of the field $\Phi$ in the Matsubara representation
is given by
\bea
\label{sigmafi}
\Sigma(\,\nu_{n}, \vec{k}\,)= 2\,g^{2}\,\int \,\frac{d^{3}\vec{p}}{(2\pi)^{3}}\,\frac{1}{\beta}\,\sum_{\omega_{m}}\,G_{\phi}
(\,\omega_{m}, \, \vec{p}\,)\,G_{\phi}(\,\omega_{m}+\nu_{n},\, \vec{p}+\vec{k}\,)\,,
\eea
where \,$\omega_m= 2\,\pi\, m/\beta$\, and \,$\nu_n=2\,\pi\, n/\beta$,\, with\, $m,\,n=0,\,\pm 1,\, \pm 2, \,\ldots$,\, are the bosonic Matsubara frequencies. The Matsubara propagators are written in the following dispersive form:
\bea
G_{\phi}(\,\omega_{m},\, \vec{p}\,) &=& \int\, dp_{0}\;\frac{\rho_{1}(\,p_{0},\,\vec{p}\,)}{p_{0}-i\, \omega_{m}}\,, \\
G_{\phi}(\,\omega_{m}+\nu_{n},\, \vec{p}+\vec{k}\,) &=& \int \,dq_{0}\;\frac{\rho_{2}(\,q_{0},\,\vec{p}+\vec{k}\,)}{q_{0}-i \,\omega_{m}-i\,\nu_{n}}\,,
\eea
where the spectral densities are
\bea
\rho_{1}(\,p_{0},\,\vec{p}\,) &=& \frac{1}{2\,\omega_1}\, [\,\delta(\,p_{0}-\omega_1\,)-\delta (\,p_{0}+\omega_1\,)\,]\,,
\quad \omega_1 = \sqrt{\vec{p}^{2}+m_\phi^{2}}\;, \\
\rho_{2}(\,q_{0},\, \vec{p}+\vec{k}\,) &=& \frac{1}{2\,\omega_2}\,[\,\delta (\,q_{0}-\omega_2\,)-\delta (\,q_{0}+\omega_2\,)\,]\,,
\quad \omega_2 = \sqrt{(\vec{p}+\vec{k})^{2}+m_\phi^{2}}\;.
\eea
This representation allows us to carry out the sum over the Matsubara frequencies $\omega_{m}$ in a rather straightforward manner
\cite{kapusta,lebellac}:
\bea \frac{1}{\beta}\,\sum_{\omega_{m}}\, \frac{1}{p_{0}-i\, \omega_{m}}\,\frac{1}{q_{0}-i \,\omega_{m}-i\,\nu_{n}}
=\frac{n_B(p_{0})-n_B(q_{0})}{q_{0}-p_0-i\,\nu_{n}}\,,
\eea
where $n_B(\om) = \frac{1}{e^{\beta\,\om} -1}$ is the Bose-Einstein distribution function. The resulting self-energy can now be written in
the dispersive form:
\bea
\label{sigdis}
\Sigma(\,\nu_n,\,\vec{k}\,) = -\frac{1}{\pi}\, \int_{-\infty}^{\infty} \,d\omega\;
\frac{\textrm{Im} \Sigma (\,\omega,\,\vec{k}\,)}{\omega-i \,\nu_n},
\eea
with $\textrm{Im} \Sigma (\,\omega,\,\vec{k}\,)$ being the imaginary part of the self-energy
\bea
\label{imsigrep}
&& \textrm{Im}\Sigma (\,\om, \,\vec{k}\,) \nonumber \\
&=& - 2\,\pi\, g^{2}\,\int\, \frac{d^{3}\vec{p}}{(2\pi)^{3}}\,\int\, dp_{0}\,dq_{0}\,
\,[\,n_B(p_{0})-n_B(q_{0})\,]\,\rho_{1}(\,p_{0},\, \vec{p}\,)\,\rho_{2}(\,q_{0},\,\vec{p}+\vec{k}\,)\,\delta(\,\omega-q_{0}+p_{0}\,)\,,
\nonumber \\
\eea

The retarded self-energy is defined by the analytic continuation:
\bea
\label{analytic}
\Sigma_\textrm{ret}(\,k_0,\,\vec{k}\,) = \Sigma(\,\nu_n=-i\, k_0 - \epsilon, \,\vec{k}\,) =-\frac{1}{\pi}\,\int_{-\infty}^{\infty}\,d\omega\,
\frac{\mathrm{Im}\Sigma_{\textrm{ret}}(\,\omega,\,\vec{k}\,)}{\omega-k_0+i\epsilon}\,.
\eea
Integrating over $dp_{0}$ and $dq_{0}$, using the identity \,$n_B(-\om)=-(\,1+n_B(\om)\,)$\, and performing the transformation $\vec{p}\rightarrow -\vec{p}-\vec{k}$ in all the integrals involving $n_B(\om_2)$, we can write
$\textrm{Im}\Sigma_\textrm{ret}(\,\om, \,\vec{k}\,)= \sigma_{0}+\sigma_{T}$ where
\bea
\sigma_{0}&=&-\frac{g^{2}}{16\, \pi^{2}}\, \textrm{sign}(\omega)\, \int \, \frac{d^{3}\vec{p}}{\omega_1\,\omega_2}\;\,
\delta(\,|\omega|-\omega_1-\omega_2\,)\,, \\
\sigma_{T}&=&-\frac{g^{2}}{8\, \pi^{2}}\, \textrm{sign}(\omega) \,\int\,
\frac{d^{3}\vec{p}}{\om_1\,\om_2}\;\, n_B(\omega_1)\; \delta(\,|\omega|-\omega_1-\omega_2\,)\,.
\eea

Obviously, $\sigma_{0}$ represents the zero-temperature contribution while $\sigma_T$ gives the finite-temperature correction. Notice that there were some possible terms involving $\delta(\,\om + \om_1 - \om_2\,)$ and $\delta(\,\om - \om_1 + \om_2\,)$ in $\textrm{Im}\Sigma_\textrm{ret}(\,\om, \,\vec{k}\,)$, but they are kinematically forbidden.  To proceed, let $\Omega=\omega_1$ and $z=\omega_2$. Then, we have
\begin{equation}
\sigma_{0}+\sigma_{T}=-\frac{g^{2}}{8\,\pi\, k}\,
\textrm{sign}(\omega)\,\int_{m_\phi}^{\infty}\, [\,1+2\,n_B(\Omega)\,] \, d\Omega\,
\int_{z^{-}}^{z^{+}}\,\delta(\,|\omega|-\Omega-z\,)\,dz\,,
\end{equation}
where $z^{\pm}$ are given by
\begin{eqnarray}
\label{z_plusminus}
z^{\pm} = \sqrt{(p\pm k)^{2}+m_{\phi}^{2}} = \sqrt{\Omega^{2}\pm 2\,k\, \sqrt{\Omega^{2}-m_{\phi}^{2}}+k^{2}}.
\end{eqnarray}
For the integral to be non-vanishing, we require that
\begin{equation}
z^{-}<z=|\omega|-\Omega<z^{+}\,.
\label{conditionI}
\end{equation}
Squaring both sides twice properly, these two inequalities can be cast into the condition $f(\Omega)<0$ where
\begin{equation}
f(\Omega)=4\,(\,|\omega|^{2}-k^{2}\,)\,\Omega^{2}-4\,|\omega|\,(\,|\omega|^{2}-k^2\,)\,\Omega+(\,|\omega|^{2}-k^2)^{2}
+4\,k^2\, m_{\phi}^{2}\,.
\end{equation}
Notice that the graph of $f(\Omega)$ against $\Omega$ represents a conic with a positive y-intercept. Solving $f(\Omega)=0$ for $\Omega$,
we obtain two solutions:
\begin{equation}
\omega^{\pm}=\frac{|\omega|\,(\,|\omega|^{2}-k^2\,)\,\pm\, k\,\sqrt{(\,|\omega|^{2}-k^2\,)^{2}-4\,(\,|\omega|^{2}-k^{2}\,)\, m_{\phi}^{2}}}{2\,(\,|\omega|^{2}-k^{2}\,)}\,.
\label{omegappm}
\end{equation}
There are two possibilities: (i) $|\omega|^{2}-k^2>0$, (ii) $k^{2}-|\omega|^{2}>0$.  For $k^{2}-|\omega|^{2}>0$, the graph with
$f(\Omega)$ against $\Omega$ shows that the condition (\ref{conditionI}) can be satisfied only if
$\Omega>\omega^{-}$ but algebraic calculations indicate that $|\omega|-\omega^{-}<0$. Hence, the condition
(\ref{conditionI}) cannot satisfied and this solution should be discarded.

For $|\omega|^{2}-k^2>0$, a detailed analysis of  $f(\Omega)$, \,$z^{\pm}$ and $|\omega|-\Omega$ as functions of $\Omega$
reveals that that condition (\ref{conditionI}) can always be satisfied as far as $\omega^{-} < \Omega<\omega^{+}$ and
$|\omega|>\sqrt{k^{2}+m_{\phi}^{2}}+m_{\phi}$. For the discriminant in $\omega^{\pm}$ to be positive, we require
$|\omega|>\sqrt{k^{2}+4\,m_{\phi}^2}$ or $|\omega|< k$. Since $\sqrt{k^{2}+m_{\phi}^{2}}+m_{\phi}> k$,
we can only choose $|\omega|>\sqrt{k^{2}+4\,m_{\phi}^2}$.

As a result, using the integration formula $\int\,\frac{d\Omega}{e^{\beta\,\Omega}-1}=\frac{1}{\beta}\,\ln\,(1-e^{-\beta\,\Omega})$, we conclude that $\textrm{Im}\Sigma_\textrm{ret}(\,\om, \,\vec{k}\,)= \sigma_{0}+\sigma_{T}$ with
\begin{eqnarray}
\sigma_{0}&=& -\frac{g^{2}}{8\, \pi\, k}\, (\,\omega^{+}-\omega^{-}\,)\; \textrm{sign}(\omega)\;
\Theta[\,|\omega|^{2}-k^{2}-4 \,m_\phi^2\,]\,,
\\
\sigma_{T}&=&-\frac{g^{2}}{4 \,\pi \,k \,\beta}\, \ln \left(\,
\frac{ 1-e^{-\beta\, \omega^{+}} }{ 1-e^{-\beta\,\omega^{-}}}\,\right)\;\textrm{sign}(\omega)
\;\Theta[\,|\omega|^{2}-k^{2}-4 \,m_\phi^2\,]\,,
\end{eqnarray}
where $\omega^{\pm}$ can now be safely simplified to become
\begin{equation}
\omega^{\pm}=\frac{|\omega|\,\pm\, k\,\sqrt{\,1-\frac{4\, m_{\phi}^{2}}{|\omega|^{2}-k^{2}}}}{2}\,.
\label{omegappmSIMPLIFIED}
\end{equation}

\subsection{Dispersion Relation: ~ Real Part of the Self-Energy}

The real part of the self-energy is given by
\bea
\textrm{Re}\,\Sigma(\nu_{n}, \vec{k})= 2\,g^{2}\,\int\, \frac{d^{3}\,\vec{p}}{(2\pi)^{3}}\,\frac{1}{\beta}\,\sum_{\omega_{m}}\,
\frac{1}{\vec{p}^{\,2}+m_{\phi}^{2}+\om_m^2}\, \frac{1}{(\vec{p}+\vec{k})^{2}+m_{\phi}^{2}+(\,\om_m+\nu_n\,)^2}\,.
\eea
To proceed, we introduce the Schwinger parameters
\bea
\frac{1}{\vec{p}^{2}+m_{\phi}^{\,2}+\om_m^2} &=& \int_{0}^{\infty}\, d\al_1 \, e^{-\al_1\, (\,\vec{p}^{2}+m_{\phi}^{2}+\om_m^2\,)} \,, \\
\frac{1}{(\vec{p}+\vec{k})^{2}+m_{\phi}^{2}+(\,\om_m+\nu_n\,)^2} &=& \int_{0}^{\infty}\, d\al_2 \,
e^{-\al_2\,  \left[\,(\vec{p}+\vec{k})^{2}+m_{\phi}^{2}+(\,\om_m+\nu_n\,)^2\,\right]}\,.
\eea
After completing squares, the $d^{3}\,\vec{p}$ integrals become Gaussian and can be easily evaluated to give
\bea
\textrm{Re}\,\Sigma(\nu_{n}, \vec{k}) &=& \frac{g^{2}}{8 \,\pi^2}\,\int_{0}^{\infty}\, d\alpha_1 \, \int_{0}^{\infty}\, d\alpha_2\,\,\frac{1}{\left(\,\alpha_1+\alpha_2\,\right)^2} \,\,\,
e^{-k_E^2 \frac{\alpha_1\,\alpha_2}{\alpha_1+\alpha_2}}\,\,\, e^{-m_\phi^2\,(\,\alpha_1+\alpha_2\,)} \nonumber \\
&& ~~~~~~~~~~~~~~~~~~~~~~~~~~~~~~~~~~~~~~~~~~~~
\vartheta\left(\,\frac{n\,\alpha_2}{\alpha_1+\alpha_2},\,\, \frac{i\,\beta^2}{4\,\pi}\frac{1}{\alpha_1+\alpha_2}\,\right)\,,
\eea
where $k_E^2 = k^2+\nu_n^2$. The Jacobi theta function $\vartheta(z,\,\tau)$ is defined as
\bea
\vartheta(z,\,\tau) =\sum_{m=-\infty}^{\infty}\,\, e^{2\,i\,\pi \,m \,z+ i \, \pi\, m^2\, \tau}\,,
\eea
and we have used the identity $\vartheta(z,\,\tau) = \left(-i\,\tau\right)^{-\frac12}\,\,e^{-i\,\pi\,z^2/\tau}\,\,\vartheta(z/\tau,\,-1/\tau)$.

Let $\alpha = \alpha_1+\alpha_2$ and $x = (\al_1-\al_2)/\al$. Then, we obtain
\bea
\label{RR1}
\textrm{Re}\,\Sigma(\nu_{n}, \vec{k}) &=& \frac{g^{2}}{16 \,\pi^2}\,\int_{0}^{\infty}\,\frac{d\,\al}{\al}\,\,
e^{\,-\frac{1}{4}\left(\,k_E^2+4\,m_\phi^2\,\right)\,\al} \nonumber \\
&& \qquad \,\,
\int_{-1}^{1}\,d\,x\,\,\,e^{\,\frac{1}{4}\,\al\,k_E^2\,x^2}\,\,\vartheta\left(\,\frac{n}{2}(1-x),\,\frac{i\,\beta^2}{4\,\pi\,\al}\,\right)\,.
\eea
To perform the integration over $dx$, we can use the formula
\bea
\int_{-1}^{1}\,d\,x\,\,e^{- A\,x + B\,x^2} =\frac{-i\,\sqrt{\pi}}{2\,\sqrt{B}}\,\,e^{-A^2/4 B}\,
\left[\, erf\left(\,i\,\frac{B+A/2}{\sqrt{B}}\,\right) + erf\left(\,i\,\frac{B-A/2}{\sqrt{B}}\,\right) \, \right]\,,~
\eea
where $erf(z) = \frac{2}{\sqrt{\pi}}\,\int_0^z \, e^{-t^2}\, dt$ is the error function. In this problem, $A = i \,\pi\, n\,m$ and $B= \frac{1}{4}\,\al\,k_E^2$. The leading contribution of the integral \eqref{RR1} comes from the region $\alpha \sim 0$. Near this region,
we have
\bea
\label{error}
e^{i\,\pi\,n\,m}\,\left[\, erf\left(\,i\,\frac{B+A/2}{\sqrt{B}}\,\right) + erf\left(\,i\,\frac{B-A/2}{\sqrt{B}}\,\right) \, \right]
\approx \frac{i\,\al^{3/2}}{2\,\pi^{5/2}}\,\frac{k_E^3}{n^2\,m^2}\,\, e^{\,\frac{1}{4}\,\al\,k_E^2}\,\, e^{-\frac{\pi^2\,n^2\,m^2}{\al\,k_E^2}}
\,.~
\eea

Meanwhile, $\textrm{Re}\,\Sigma(\nu_{n}, \vec{k})$ can be written as
\bea
\textrm{Re}\,\Sigma(\nu_{n}, \vec{k}) = \sum_{m=-\infty}^{+\infty}\,I_m = I_0 + \sum_{m=-\infty}^{+\infty}\,I_{m\neq 0}\,.
\eea
The quantity $I_0$ corresponds to the zero-temperature contribution and we assume that it has already been combined with the bare
mass-squared of $\Phi$ to give $M_\Phi^2$. Therefore, the mass-squared of $\Phi$ at finite-temperature $T$ takes the form
$M_\Phi^2(T) = M_\Phi^2 + \Delta M_\Phi^2(T)$ with $\Delta M_\Phi^2(T)$ being the finite-temperature corrections:
\bea
\Delta M_\Phi^2(T) = \sum_{m=-\infty}^{+\infty}\,I_{m\neq 0}\,.
\eea

Upon some simplifications, we get
\bea
\label{dalpha}
I_{m\neq 0} \approx \frac{g^{2}}{32 \,\pi^4}\,\frac{k_E^2}{n^2\,m^2}\,\int_{0}^{\infty}\,d\,\al\,\,
e^{-\alpha\,m_\phi^2 \,-\,\frac{1}{\al}\,\left(\frac{\beta^2\,m^2}{4}\right) }\,,
\eea
which is an even function of $m$ and so $\Delta M_\Phi^2(T) = 2\,\sum_{m=1}^{+\infty}\,I_{m}$. We can perform the remaining integration
using the identity
\bea
\int_{0}^{\infty}\,d\,\al\,\, e^{-\alpha\,C \,-\,\frac{1}{\al}\,D } =2\,\,\sqrt{\frac{D}{C}}\,\,K_1(\,2\,\sqrt{C\,D}\,)\,,
\eea
where $K_1(z)$ is the modified Bessel function of second kind.

Applying the analytic continuation:\, $\nu_n = \frac{2\,\pi\,n}{\beta} \rightarrow -i\,\omega - \epsilon$, we find
\bea
\Delta M_\Phi^2(T) \approx \frac{g^{2}}{4 \,\pi^2}\,\,\frac{\omega^2- k^2}{\beta \,\omega^2 \, m_\phi}\,\,
\sum_{m=1}^{+\infty}\,\frac{1}{m} \, K_1(m\,\beta \, m_\phi)\,.
\eea
Since $K_1(z) \sim \sqrt{\frac{\pi}{2\,z}}\,e^{-z}$ for $z \gg 1$, it is obvious that $\frac{1}{m} \, K_1(m\,\beta \, m_\phi)$ will be
exponentially suppressed if $m\,\beta\,m_\phi \gg 1$. On the other hand, $K_1(z) \sim \frac{1}{z}$ for $z \ll 1$. Thus, the dominant
contribution of $\frac{1}{m} \, K_1(m\,\beta \, m_\phi)$ goes like $\frac{1}{m}\, \frac{1}{m\,\beta\,m_\phi}$.
Using $\sum_{m=1}^{\infty}\,\frac{1}{m^2}=\frac{\pi^2}{6}$, we obtain
\bea
\Delta M_\Phi^2(T) \approx \frac{g^{2}}{24}\,\,\frac{\omega^2- k^2}{\omega^2}\,\frac{T^2}{m_\phi^2}\,.
\eea

\section{~~ Pseudoscalar $\Phi \rightarrow f\, \bar{f}$}

The one-loop self-energy of the field $\Phi$ in the Matsubara representation is given by
\bea
\label{sigmafiFermion}
\Sigma(\,\nu_{n}, \vec{k}\,)= -\lambda^{2}\,\int\,
\,\frac{d^{3}\vec{p}}{(2\pi)^{3}}\,\frac{1}{\beta}\,\sum_{\omega_{m}}\,\textrm{Tr}\,\left[\,G_f(\,\omega_{m},\,\vec{p}\,)\,\gamma^5\,
G_{\bar{f}}(\,\omega_{m}+\nu_{n},\,\vec{p}+\vec{k}\,)\,\gamma^5\,\right]\,,
\eea
where \,$\omega_m= 2\,\pi\, (m+\frac12)/\beta$\, and \,$\nu_n=2\,\pi\, (n+\frac12)/\beta$,\, with\, $m\,,n=0,\,\pm 1,\, \pm 2, \,\ldots$,\,
are the fermionic Matsubara frequencies. It is convenient to write the Matsubara propagators in the dispersive form:
\begin{eqnarray}
G_f(\,\omega_{m}, \,\vec{p}\,)&=& \int\, dp_{0}\;\frac{\rho_{1}(\,p_{0},\,\vec{p}\,)}{p_{0}-i\, \omega_{m}}\,, \\
G_{\bar{f}}(\,\omega_{m}+\nu_{n},\,\vec{p}+\vec{k}\,) &=& \int\,  dq_{0}\;\frac{\rho_{2}(\,q_{0}, \,\vec{p}+\vec{k}\,)}{q_{0}-i\,
\omega_{m}-i\,\nu_{n}}\,,\\
\rho_{1}(\,p_{0},\,\vec{p}\,)&=& \frac{\gamma^0\,p_0-\vec{\gamma}\cdot \vec{p}+m_f}{2\,\omega_1} \,
[\,\delta(\,p_{0}-\omega_1)-\delta(\,p_{0}+\omega_1\,)\,]\,, \\
\rho_{2}(\,q_{0},\,\vec{p}+\vec{k}\,) &=& \frac{\gamma^0\,q_0-\vec{\gamma}\cdot (\vec{p}+\vec{k})+m_f}{2\,\omega_2}
\,[\,\delta(\,q_{0}-\omega_2\,)-\delta(\,q_{0}+\omega_2\,)\,]\,,\\
\omega_1&=& \sqrt{\vec{p}^{2}+m_f^2}\,,
\quad \omega_2 = \sqrt{(\vec{p}+\vec{k})^{2}+m_f^2}\,.
\end{eqnarray}
This representation allows us to carry out the sum over the Matsubara frequencies $\omega_{m}$ in a rather straightforward manner
\cite{kapusta,lebellac}:
\bea
\frac{1}{\beta}\,\sum_{\omega_{m}}\, \frac{1}{p_{0}-i\, \omega_{m}}\,\frac{1}{q_{0}-i\, \omega_{m}-i\,\nu_{n}} =-\frac{n_F(p_{0})-n_F(q_{0})}{q_{0}-p_0-i\,\nu_{n}} \,,
\eea
where $n_F(\om) = \frac{1}{e^{\beta\,\om} +1}$ is the Fermi-Dirac distribution function. The self-energy can be written in the dispersive form:
\begin{equation}
\label{sigdisFermion}
\Sigma(\,\nu_n,\,\vec{k}\,) = -\frac{1}{\pi}\, \int_{-\infty}^{\infty}\,
d\omega\; \frac{\textrm{Im} \Sigma (\,\omega,\,\vec{k}\,)}{\omega-i\, \nu_n}\,,
\end{equation}
where $\textrm{Im}\Sigma(\,\omega,\,\vec{k}\,)$ is the imaginary part of the self-energy given by
\begin{eqnarray}
\label{imsigrepFermion}
&&\textrm{Im}\Sigma(\,\omega,\,\vec{k}\,)  \nonumber \\
&=& \pi\,\la^{2} \,\int \,\frac{d^{3}\vec{p}}{(2\pi)^{3}}\,\int\, dp_{0}\, dq_{0}\,[\,n_F(p_{0})-n_F(q_{0})\,]\;
\textrm{Tr}\left(\,\rho_{1}(\,p_{0},\vec{p}\,)\,\gamma^5\,\rho_{2}(\,q_{0},\,\vec{p}+\vec{k}\,)\,\gamma^5\,\right)\,
\delta(\,\omega-q_{0}+p_{0}\,)\,, \nonumber \\
\end{eqnarray}
We can then proceed by using $\textrm{Tr}(1)=4$ and
$\textrm{Tr}(\gamma^{\mu}\,\gamma^{\nu})=4\,g^{\mu\nu}$, giving
\bea
&& \textrm{Tr}\left[\,\left(\,\gamma^0\,p_0-\vec{\gamma}\cdot\vec{p}+m\,\right)\,\gamma^5\,
\left(\,\gamma^0\,q_0-\vec{\gamma}\cdot (\vec{p}+\vec{k})+m\,\right)\,\gamma^5 \,\right] \nonumber \\
&=& - 4 \left(\,p_0\,q_0-\vec{p}\cdot(\vec{p}+\vec{k})-  m^2\,\right)\,.
\eea

The retarded self-energy is defined by the same analytic continuation as in Eq. \eqref{analytic}. Similarly,
integrating over $dp_{0}$ and $dq_{0}$, using the identity \,$n_F(-\om)=1-n_F(\om)$\, and performing the transformation $\vec{p}\rightarrow -\vec{p}-\vec{k}$ in all the integrals involving $n_F(\om_2)$, we can write
$\textrm{Im}\Sigma_\textrm{ret}(\,\om, \,\vec{k}\,)= \sigma_{0}+\sigma_{T}$ where
\bea
\sigma_{0}&=&-\frac{\la^{2}}{8\, \pi^{2}}\, \textrm{sign}(\omega)\, \int \, \frac{d^{3}\vec{p}}{\omega_1\,\omega_2}\;\,
\left(\,\om_1\,\om_2 + \vec{p}\cdot(\vec{p}+\vec{k}) + m_f^2 \,\right)\,\delta(\,|\omega|-\omega_1-\omega_2\,)\,, \\
\sigma_{T}&=& \frac{\la^{2}}{4\, \pi^{2}}\, \textrm{sign}(\omega) \,\int\,
\frac{d^{3}\vec{p}}{\om_1\,\om_2}\;\, n_F(\omega_1)\,\left(\,\om_1\,\om_2 + \vec{p}\cdot(\vec{p}+\vec{k}) + m_f^2 \,\right)\, \delta(\,|\omega|-\omega_1-\omega_2\,)\,.
\eea

Again, $\sigma_{0}$ represents the zero-temperature contribution while $\sigma_T$ gives the finite-temperature correction. Also, there were some possible terms in $\textrm{Im}\Sigma_\textrm{ret}(\,\om, \,\vec{k}\,)$ involving $\delta(\,\om + \om_1 - \om_2\,)$ and $\delta(\,\om - \om_1 + \om_2\,)$ which are kinematically forbidden. To proceed, we again let $\Omega=\omega_1$ and $z=\omega_2$. Then, we have\,
$\delta(\,|\omega|-\omega_1-\omega_2\,) = \delta(\,|\omega|-\Omega- z\,)$, and we can make the following simplification:
\bea
\om_1\,\om_2 + \vec{p}\cdot(\vec{p}+\vec{k}) + m_f^2 = \frac{\left(\,\Omega+z\,\right)^2-k^2}{2} = \frac{|\om|^2-k^2}{2}\,,
\eea
using the constraint\, $\delta(\,|\omega|-\Omega- z\,)$. This leads to
\begin{equation}
\sigma_{0}+\sigma_{T}=-\frac{\la^{2}}{8\,\pi\, k}\,\left(\,|\om|^2-k^2\,\right)\,
\textrm{sign}(\omega)\,\int_{m_f}^{\infty}\, [\,1-2\,n_F(\Omega)\,] \, d\Omega\,
\int_{z^{-}}^{z^{+}}\,\delta(\,|\omega|-\Omega-z\,)\,dz\,,
\end{equation}
where $z^{\pm}$ are given by Eq. \eqref{z_plusminus} with $m_\phi$ replaced by $m_f$. We can then follow the similar kinematical arguments in Appendix A to facilitate the integrations over both of $d\Omega$ and $dz$.

As a result, using the integration formula $\int\,\frac{d\Omega}{e^{\beta\,\Omega}+1}=-\frac{1}{\beta}\,\ln\,(1+e^{-\beta\,\Omega})$, we conclude that $\textrm{Im}\Sigma_\textrm{ret}(\,\om, \,\vec{k}\,)= \sigma_{0}+\sigma_{T}$ with
\begin{eqnarray}
\sigma_{0}&=& -\frac{\la^{2}}{8\, \pi\, k} \,\left(\,|\om|^2-k^2\,\right)\,(\,E^{+}-E^{-}\,)\;
\textrm{sign}(\omega)\; \Theta[\,|\omega|^{2}-k^{2}-4 \,m_f^2\,]\,,
\\
\sigma_{T}&=&-\frac{\la^{2}}{4 \,\pi \,k \,\beta}\,\left(\,|\om|^2-k^2\,\right)\, \ln \left(\,
\frac{ 1+e^{-\beta\, E^{+}} }{ 1+e^{-\beta\,E^{-}}}\,\right)\; \textrm{sign}(\omega)
\;\Theta[\,|\omega|^{2}-k^{2}-4 \,m_f^2\,]\,,
\end{eqnarray}
where $E^{\pm}$ is given by
\begin{equation}
E^{\pm}=\frac{|\omega|\,\pm\, k\,\sqrt{\,1-\frac{4\, m_{f}^{2}}{|\omega|^{2}-k^{2}}}}{2}\,.
\label{omegappmFermion}
\end{equation}

\section{~~ Scalar $\Phi \rightarrow f\, \bar{f}$}

Similar to Appendix B, the one-loop self-energy of the field $\Phi$ in the Matsubara representation is given by
\bea
\label{sigmafiFermion1}
\Sigma(\,\nu_{n}, \vec{k}\,)= y^{2}\,\int\,
\,\frac{d^{3}\vec{p}}{(2\pi)^{3}}\,\frac{1}{\beta}\,\sum_{\omega_{m}}\,\textrm{Tr}\,\left[\,G_f(\,\omega_{m},\,\vec{p}\,)\,
G_{\bar{f}}(\,\omega_{m}+\nu_{n},\,\vec{p}+\vec{k}\,)\,\right]\,,
\eea
where \,$\omega_m= 2\,\pi\, (m+\frac12)/\beta$\, and \,$\nu_n=2\,\pi\, (n+\frac12)/\beta$,\, with\, $m\,,n=0,\,\pm 1,\, \pm 2, \,\ldots$,\,
are the fermionic Matsubara frequencies.

Following the similar steps and tricks as in Appendix B, we obtain
$\textrm{Im}\Sigma_\textrm{ret}(\,\om, \,\vec{k}\,)= \sigma_{0}+\sigma_{T}$ with
\begin{eqnarray}
\sigma_{0}&=& -\frac{\alpha^{2}}{8\, \pi\, k} \,\left(\,|\om|^2-k^2-4\,m_f^2\,\right)\,
(\,E^{+}-E^{-}\,)\;
\textrm{sign}(\omega)\; \Theta[\,|\omega|^{2}-k^{2}-4 \,m_f^2\,]\,,
\\
\sigma_{T}&=&-\frac{\alpha^{2}}{4 \,\pi \,k \,\beta}\,\left(\,|\om|^2-k^2-4\,m_f^2\,\right)\, \ln \left(\,
\frac{ 1+e^{-\beta\, E^{+}} }{ 1+e^{-\beta\,E^{-}}}\,\right)\; \textrm{sign}(\omega)
\;\Theta[\,|\omega|^{2}-k^{2}-4 \,m_f^2\,]\,,
\end{eqnarray}
where $E^{\pm}$ is given by \eqref{omegappmFermion}.

{}

\end{document}